\documentclass[aps,prd,twocolumn,superscriptaddress]{revtex4-1}
\usepackage{graphicx}
\begin{document}

%\preprint{}

\title{Study of muons from ultra-high energy cosmic ray air showers measured with the Telescope Array experiment}


%\author{}
%\thanks{}
%\affiliation{}

\author{R.U.~Abbasi}
\affiliation{High Energy Astrophysics Institute and Department of Physics and Astronomy, University of Utah, Salt Lake City, Utah, USA}
\author{M.~Abe}
\affiliation{The Graduate School of Science and Engineering, Saitama University, Saitama, Saitama, Japan}
\author{T.~Abu-Zayyad}
\affiliation{High Energy Astrophysics Institute and Department of Physics and Astronomy, University of Utah, Salt Lake City, Utah, USA}
\author{M.~Allen}
\affiliation{High Energy Astrophysics Institute and Department of Physics and Astronomy, University of Utah, Salt Lake City, Utah, USA}
\author{R.~Azuma}
\affiliation{Graduate School of Science and Engineering, Tokyo Institute of Technology, Meguro, Tokyo, Japan}
\author{E.~Barcikowski}
\affiliation{High Energy Astrophysics Institute and Department of Physics and Astronomy, University of Utah, Salt Lake City, Utah, USA}
\author{J.W.~Belz}
\affiliation{High Energy Astrophysics Institute and Department of Physics and Astronomy, University of Utah, Salt Lake City, Utah, USA}
\author{D.R.~Bergman}
\affiliation{High Energy Astrophysics Institute and Department of Physics and Astronomy, University of Utah, Salt Lake City, Utah, USA}
\author{S.A.~Blake}
\affiliation{High Energy Astrophysics Institute and Department of Physics and Astronomy, University of Utah, Salt Lake City, Utah, USA}
\author{R.~Cady}
\affiliation{High Energy Astrophysics Institute and Department of Physics and Astronomy, University of Utah, Salt Lake City, Utah, USA}
% ...

\collaboration{Telescope Array Collaboration}
\noaffiliation

\date{\today}

\begin{abstract}
One of the uncertainties in interpretation of ultra-high energy cosmic ray (UHECR) data comes from the hadronic interaction models used for air shower Monte Carlo (MC) simulations.
The number of muons observed at the ground from UHECR-induced air showers is expected to depend upon the hadronic interaction model.
One may therefore test the hadronic interaction models by comparing the measured number of muons with the MC prediction.
In this paper, we present the results of studies of muon densities in UHE extensive air showers obtained by analyzing the signal of surface detector stations which should have high $\it{muon \, purity}$. 
The muon purity of a station will depend on both the inclination of the shower and the relative position of the station.
In 7~years' data from the Telescope Array experiment, we find that the number of particles observed for signals with an expected muon purity of $\sim$65\% at a lateral distance of 2000 m from the shower core is $1.72 \pm 0.10{\rm (stat.)} \pm 0.37 {\rm (syst.)}$ times larger than the MC prediction value using the QGSJET I$\hspace{-.1em}$I-03 model for proton-induced showers.
A similar effect is also seen in comparisons with other hadronic models such as QGSJET I$\hspace{-.1em}$I-04, which shows a $1.67 \pm 0.10
\pm 0.36$ excess.
We also studied the dependence of these excesses on lateral distances and found a slower decrease of the lateral distribution of muons in the data as compared to the MC, causing larger discrepancy at larger lateral distances.
\end{abstract}


\maketitle

\section{Introduction\label{sec:1}} 

The origin of ultra-high energy cosmic rays (UHECRs) has been a long-standing mystery of astrophysics.
The Telescope Array (TA) experiment~\cite{TA} in Utah, USA is the largest experiment in the northern hemisphere observing UHECRs. 
It aims to reveal the origin of UHECRs by studying the energy spectrum, mass composition and anisotropy of cosmic rays.
When a UHECR enters the atmosphere, it interacts with atmospheric nuclei and generates the particle cascade, which is called an air shower.
The information of primary cosmic rays is estimated from observed signals of air shower particles and the air shower Monte Carlo (MC) simulation.

UHECR air showers are not fully understood.
At present, the maximum target-frame energy of hadronic interactions accessible at accelerators is 10$^{17}$~eV at the CERN LHC.
The MC for cosmic rays in the energies above $10^{18}$~eV uses the extrapolated values of the parameters of hadronic interactions, such as cross section and multiplicity.
The values of these parameters differ between hadronic interaction models, due to the uncertainty of modeling pion or kaon generation at the early age of the air shower development.
Thus inferences of UHECR composition from air shower measurements are model-dependent~\cite{Xmax_stereo, Xmax_mono},
which is important to understand the origin of UHECRs because cosmic rays are deflected in the galactic and extragalactic magnetic fields.

In addition to that, HiRes/MIA experiment reported a deficit in the number of muons from MC air showers compared with experimental data for $E \gtrsim 10^{17}$~eV~\cite{hir}.
The Yakutsk experiment also indicated lower simulated muon densities than those observed for $E \gtrsim 10^{19}$~eV~\cite{yak}.
The Pierre Auger Observatory, which is located in Mendoza, Argentina, reported~\cite{Au_Nmu1} a model-dependent deficit of muons in simulations of 30--80\% relative to the data, $10^{19}$~eV.
The Auger group also reported that the observed hadronic signal in UHECR air showers is $1.61 \pm 0.21$ $(1.33 \pm 0.16)$ times larger than the post-LHC MC prediction values for QGSJET I$\hspace{-.1em}$I-04~\cite{QII-04} (EPOS-LHC~\cite{EP-LHC}), including statistical and systematic errors~\cite{Au_Nmu2}.
For $E \lesssim 10^{17}$~eV, The KASCADE-Grande experiment~\cite{kas} and an atmospheric muon study~\cite{atm} reported the discrepancy between experimental data and the air shower MC models, whereas the Icetop preliminary result~\cite{ice} and the EAS-MSU array~\cite{eas} showed no excess of muons in their data.

The analysis of air shower components provides the information to obtain a realistic air shower model.
The number of muons from a UHECR on the ground depends on the mass composition of primary cosmic rays.
The MC prediction depends also on hadronic interaction models since it has information about the shower development at an early stage.
One may test the hadronic interaction models by comparing the measured number of muons with the MC prediction.
Furthermore, the lateral distribution of muons -- which was not analyzed in the previous studies~\cite{hir, yak, Au_Nmu1, Au_Nmu2} -- contains information about the hadronic interaction. 
In this work, we study the difference between the number of muons in experimental data and the MC. 
We also study the difference as a function of lateral distance from the shower core.
For that, an analysis for muons from UHECR air showers with the TA SD was developed.

\section{Telescope Array experiment} \label{sec:2}

\subsection{Detector and data taking}
{
The TA experiment is designed to measure air shower particles on the ground with the SDs and fluorescence light induced by the air shower with the fluorescence detectors (FDs).
The TA SD array consists of 507 scintillation counters, placed on a square grid with 1.2 km spacing, covering 700 km$^2$ \cite{SD2012}.
Each TA SD is composed of two layers of plastic scintillator with two photomultiplier tubes (PMTs), one for each layer. 
It has an area of 3 m$^2$ and each layer has 1.2 cm thickness.
The scintillators and PMTs are contained in a stainless steel box which is mounted under a 1.2 mm thick iron roof to protect the detector from large temperature variations.
The SD station is solar-powered and data are collected by a Wireless Local-Area Network (WLAN) system.
The station measures air shower particles by collecting photons generated in scintillators through wavelength shifting fibers and detecting them with PMTs.
The signals from PMTs are digitized by a commercial 12-bit Flash Analog-Digital Converter (FADC) with a 50 MHz sampling rate on the CPU board.
The SD array trigger is created when at least three adjacent counters detect energy deposits equivalent to greater than the equivalent of three minimum ionizing particles (MIPs) within 8 microseconds.  
The readout system then records SD signals equivalent to $\gtrsim 0.3$~MIP detected within $\pm{32}$ microseconds of the trigger time. 
The trigger efficiency is greater than 97\% for primary particles with energies above 10$^{19}$~eV \cite{SD2012}.
The calibration is performed every 10 minutes by monitoring histograms of signals from single atmospheric muons and comparing them with simulated distributions of energy deposition \cite{SD2012}, thus determining the correlation between the FADC values and the energy deposition.

The three TA FD stations are located around the SD array and view the sky above the array \cite {FD_BRLR}.
The stations consists of 38 telescopes with spherical mirrors.
The fluorescence light from air showers is collected by the mirror and detected by PMTs through a UV band-pass filter.
The signals from each PMT are digitized by a 12-bit FADC with a 40 MHz sampling rate.
The trigger electronics select a track pattern of triggered PMTs in real time and record air shower tracks.
}

\subsection{Event reconstruction}
{
The TA SD event reconstruction consists of the following steps~\cite {DI}: 
First, SD signals that are related to air shower events are selected by determining a cluster which is contiguous in space and time.
Signals less than about 1.4 VEM (Vertical Equivalent Muon) are excluded from the cluster.
This process reduces background signals from the random atmospheric muons, which occur uniformly in space and time at a rate of 0.05 per station within one event time period ($\pm{32}$ microseconds).
Second, a time fit of shower arrivals at the SDs is performed
to determine the geometry of cosmic ray air showers.
Third, the lateral distribution of charged particle densities at the SDs is fit using the AGASA lateral distribution function \cite{lat1,lat2} to determine $S$800, the density of shower particles at a lateral distance of 800 m from the air shower axis.
The energy of the cosmic ray is estimated by using a look-up table in $S$800 and the shower zenith angle.
The table is obtained by a large statistics MC simulation using CORSIKA \cite {CORSIKA} and the QGSJET I$\hspace{-.1em}$I-03 hadronic model.
Finally, the reconstructed energy is scaled to the energy measured by the TA FD, 
which is determined using calorimetric detection of an air shower energy deposition in the atmosphere~\cite {FD_BRLR, Xmax_BRLRhybrid2, BRLRhybrid3} with less hadronic interaction dependence than the SD.
The energy and angular resolutions for a primary energy within $10^{18.5}$~eV~$< E <$~$10^{19.0}$~eV are 29\% and 2.1$^\circ$, respectively~\cite{DI_2017}. Those for energies above $10^{19}$~eV are 19\% and 1.4$^\circ$, respectively.
The resolution of the distance from a shower axis is about 70 - 80 m within $10^{18.8}$~eV~$< E <$~$10^{19.2}$~eV, which is the energy range analyzed in this work.

\section{Monte Carlo simulation} \label{sec:3}

We use CORSIKA v6.960, and QGSJET I$\hspace{-.1em}$I-03 as a reference model for high energy hadronic interactions.
The MC for other models are also generated using the same MC procedure.
We also use FLUKA2008.3c \cite{flu1, flu2} to model low energy hadronic interactions and EGS4 \cite{egs4} to model electromagnetic interactions.
The thinning \cite{thin} and ''de-thinning'' \cite{dethin}  techniques are used to reduce the computation time.
The detector simulation is performed by using Geant4 \cite{Geant4} toolkit.

The simulated cosmic ray energies range from $10^{16.55}$ to $10^{20.55}$~eV.
The simulated zenith angle is isotropically distributed from 0$^{\circ}$ to 60$^{\circ}$.
The azimuth angle and core position are randomly distributed within the SD array.
The same reconstruction procedure as experimental data is applied for the MC dataset.
We sampled simulated events so that the energy distribution follows the spectrum measured by the HiRes experiment \cite{hires}.
The distributions of the reconstructed shower parameters, such as energy and zenith angle distributions, are in good agreement between the experimental data and the MC \cite{DI}.

\section{Analysis procedure} \label{sec:4}

\subsection{Dataset}
We use the TA SD 7 years' dataset recorded from 11~May~2008 through 11~May~2015, and the events reconstructed by the same method as the TA spectrum analysis \cite {TAspe} 
with an energy range $10^{18.8}$~eV~$< E <$~$10^{19.2}$~eV.
In this energy range, the mass composition of the primary cosmic rays is consistent with a light component within statistical and systematic errors as determined by $X_{\rm{max}}$ measurement using the TA FD~\cite{Xmax_stereo, Xmax_mono, Xmax_MDhybrid2, Xmax_BRLRhybrid2}, where $X_{\rm{max}}$ is the depth in the atmosphere of air shower maximum, thus we use the MC for proton primaries.
We used the energy scale corrected by the FD (reconstructed energy scale) for the experimental data and the scale not corrected by the FD (thrown energy scale) for the MC, since the difference of an energy scale corresponds to the difference of the number of particles in SD signals at the same energy.
The experimental data are compared with the MC using the hadronic models QGSJET I$\hspace{-.1em}$I-03, QGSJET I$\hspace{-.1em}$I-04, Epos 1.99 \cite{EP199} and Sibyll 2.1 \cite{Sib2.1}.

\subsection{Analysis framework}
The TA SD, made of plastic scintillators \cite{SD2012}, is sensitive to the electromagnetic (EM) component (electrons and gammas) that are the predominant part of secondary particles from the air showers.
The conversion rate of gammas to electrons in the TA SD is $\sim$20\% at 1 GeV.
To increase the ratio of muons in SD signals, we used the following analysis approach.
Firstly we define the condition of the high $\it{muon\, purity}$ using the MC.
Then we compare the observed signal size from air shower particles with the MC prediction under the high muon purity condition.

The secondary particles generated in the atmosphere are attenuated by the interaction with atmospheric particles and they decay before they reach the ground. 
The EM components experience greater attenuation than muons over the same path length, because the EM components largely lose their energy by the pair production and bremsstrahlung in the shower development but muons can penetrate the atmosphere down to the ground before their decay.
Hence the ratio of the energy deposit of air shower muons to that of all particles, which consist of air shower and background components, in SD signals (hereafter this ratio is described as the muon purity)  is expected to be larger for SDs more distant from secondary particle generation points on the shower axis.
We classify the detector hits in the air shower events of the dataset using $\theta$ (the zenith angle), 
$\phi$ (the azimuth angle relative to the shower arrival direction projected onto the ground), and $R$ (the distance from a shower axis).
The geometry definition is described in Figure \ref {fig:ana_geom}.
When $\theta, |\phi|$ or $R$ values become large, the path length of air shower particles 
increases, then the muon purity in SD signals is expected to be high. 

\begin{figure}
\includegraphics[width=0.5\textwidth] {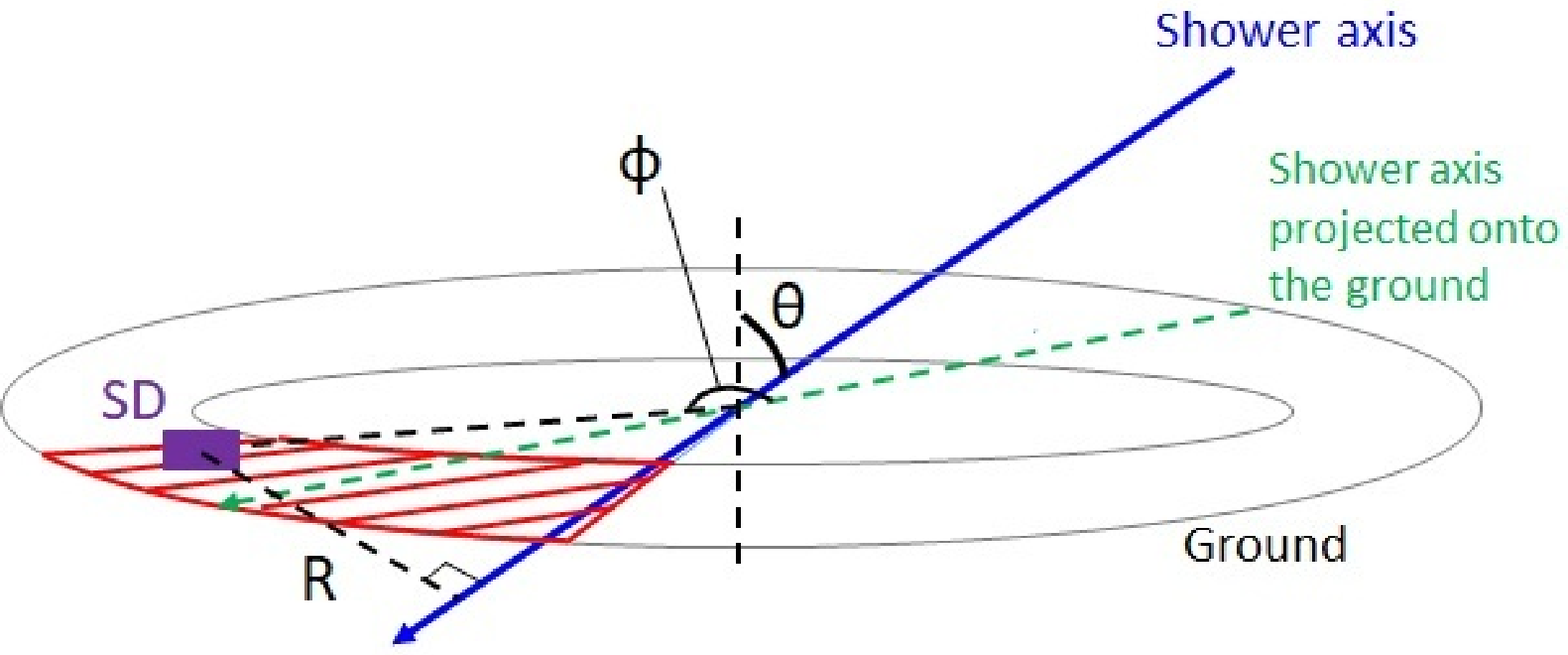}
   \includegraphics[width=0.33\textwidth] {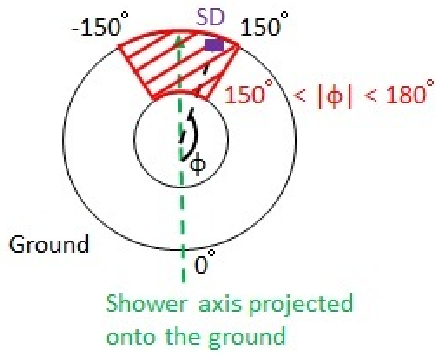}
\caption{(color online).
(top) Geometry definition of the muon analysis. 
An SD location on the ground is selected by $\phi$ and $R$ to reduce the EM background. 
The muon purity in the SD signal is calculated in each ($\phi, R$) bin. 
There are six bins for $\phi$ and 18 bins for $R$ from 500 to 4500 m.
The red shaded region in the figure shows the bin for $150^{\circ} < |\phi| < 180^{\circ}$ where the distance from the particle generation points on the shower axis is relatively larger than other $\phi$ bins, which is expected to be the less EM background bin. 
(bottom) Top view for the $\phi$ definition. 
\label{fig:ana_geom}}
\end{figure}

The integrated FADC is calculated for each SD participating in the event.
The FADC count, converted to VEM units, is entered in the histogram of the corresponding ($\theta, \phi, R$) bin.
Figure \ref{fig:wf} and \ref{fig:cLDF_1} show sample waveforms and histograms for each particle type.
One detector signal corresponds to one entry in the histogram.
An SD which has no signal is assigned to the 0~VEM bin of the histogram.
Since the cut on hit signals less than about 1.4 VEM is applied to the total (black), there are remaining entries below the signal size for other components (other colors).
Figure \ref{fig:cLDF_2} shows the lateral distributions of SD signals and the muon purity.
The muon purity is mainly 60 - 70\% on the high muon purity condition ($30^{\circ} < \theta < 45^{\circ}, 150^{\circ} < |\phi| < 180^{\circ},$ 2000 m $< R <$ 4000 m).
We use these conditions to select high muon purity events for the comparison of the data with the MC.

\begin{figure}
   \includegraphics[width=0.45\textwidth] {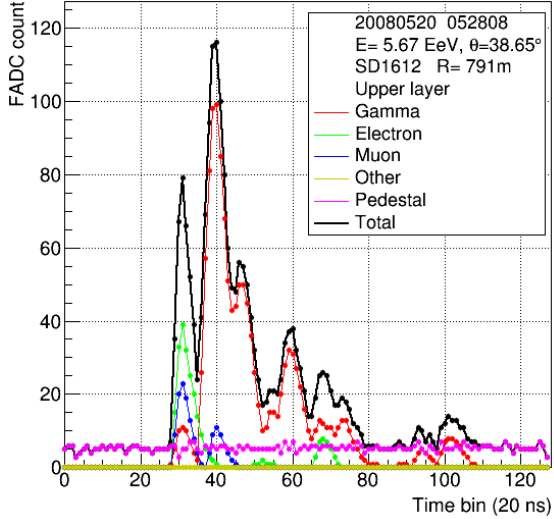}
   \includegraphics[width=0.45\textwidth] {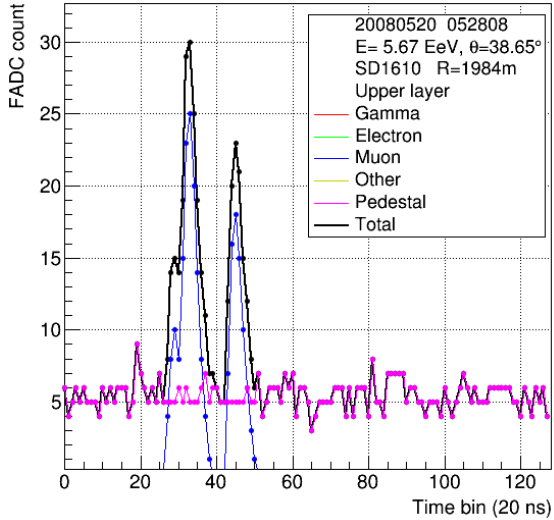}
\caption{(color online).
(top) Sample event waveforms of different particle types in an SD (3 m$^2$ in area) at $R \simeq 800$~m made by the MC.
The red, green, blue, yellow, magenta and black represent gamma, electron, muon, other shower components, atmospheric muon background and the total of them, respectively.
(bottom) Same as top figure, but at $R \simeq 2000$ m.
The components except muons and pedestals have 0 FADC values in this sample.
\label{fig:wf}}
\end{figure}

\begin{figure}
   \includegraphics[width=0.45\textwidth] {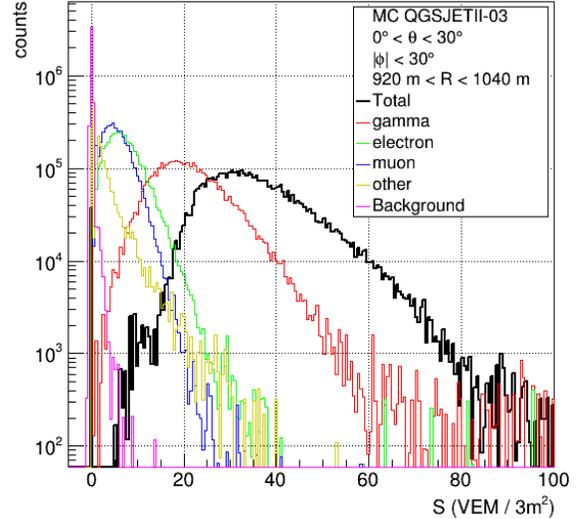}
   \includegraphics[width=0.45\textwidth] {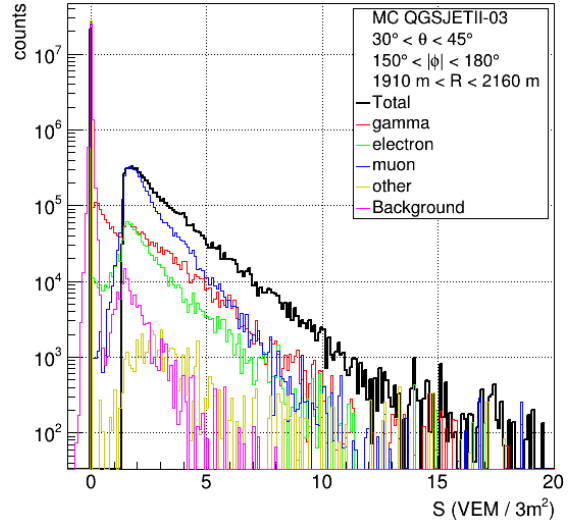}
\caption{(color online).
(top) Histograms of the signal size (denoted as $S$ in the figure) of different particle types in an SD made by the MC for $10^{18.8}~{\rm eV} < E < 10^{19.2}~{\rm eV}, 0^{\circ} < \theta < 30^{\circ}, |\phi| < 30^{\circ}$, 920 m $< R <$ 1040 m.
The red, green, blue, yellow, magenta and black represent gamma, electron, muon, other shower components, atmospheric muon background and the total of them, respectively.
(bottom) Same as top figure, but for $30^{\circ} < \theta < 45^{\circ}, 150^{\circ} < |\phi| < 180^{\circ}$, 1910 m $< R <$ 2160 m.
\label{fig:cLDF_1}}
\end{figure}

\begin{figure}
   \includegraphics[width=0.45\textwidth] {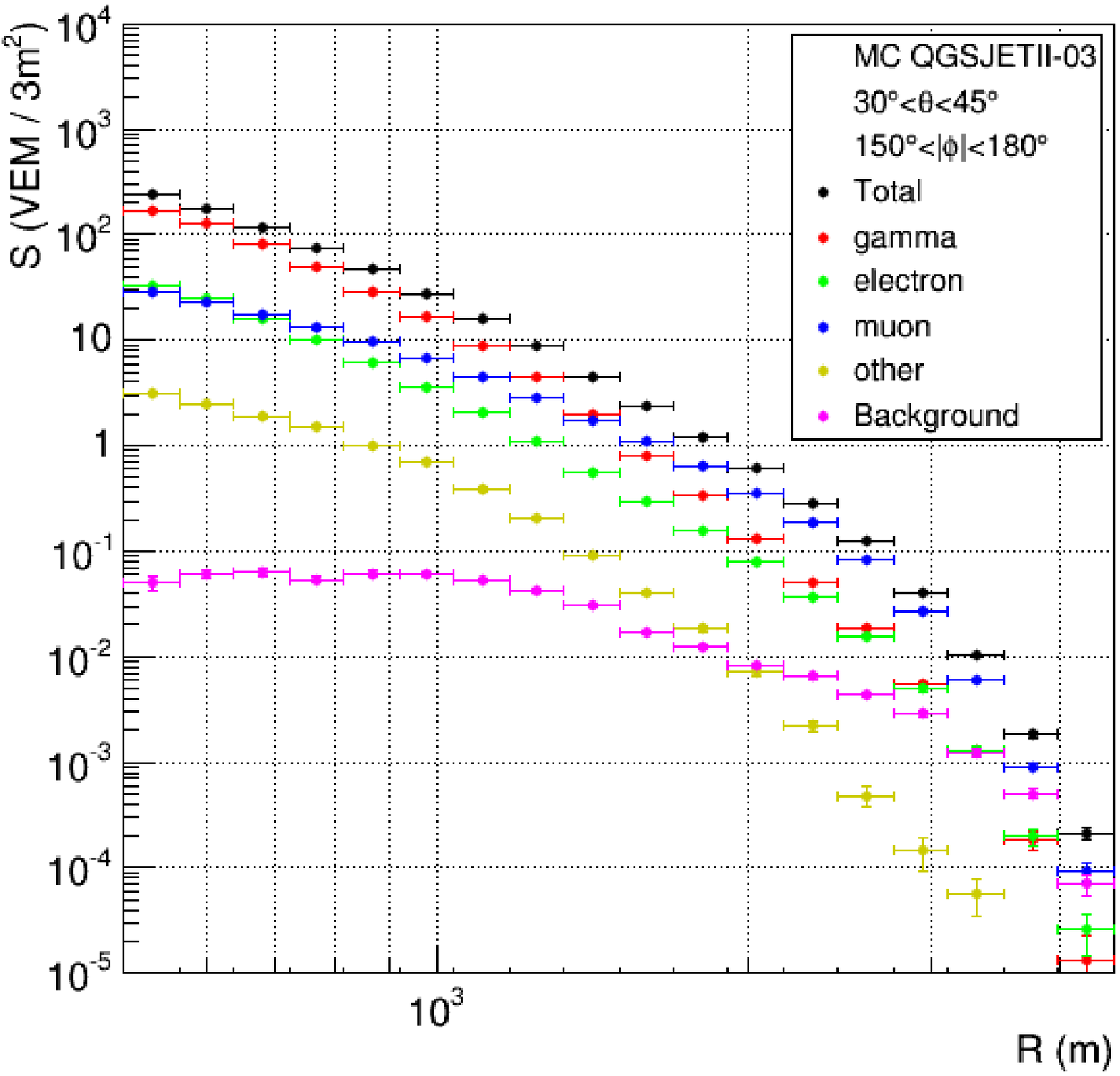}
   \includegraphics[width=0.45\textwidth] {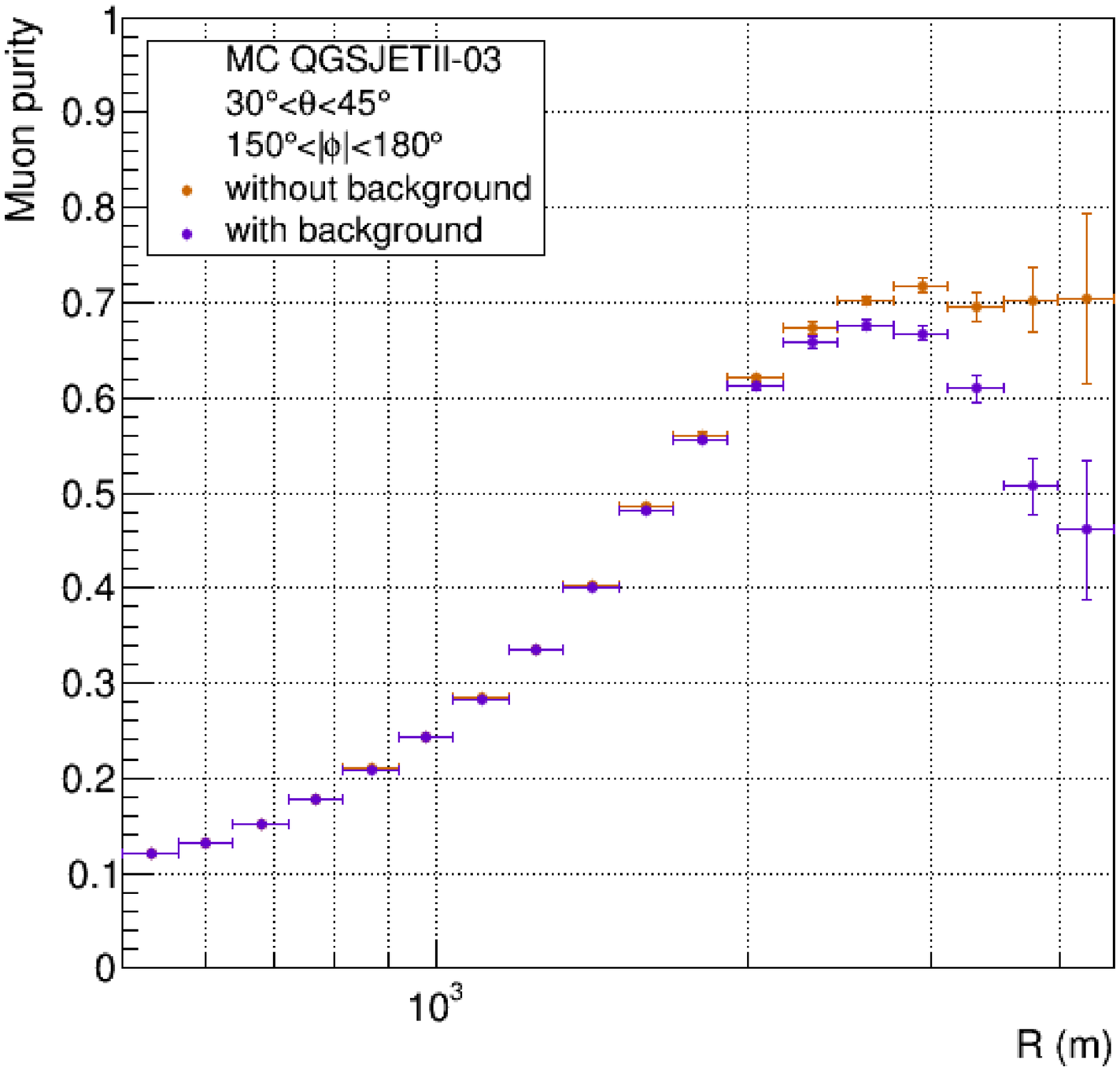}
\caption{(color online).
(top) Lateral distributions of the air shower average signal of the MC with QGSJET I$\hspace{-.1em}$I-03 for $30^{\circ} < \theta < 45^{\circ}, 150^{\circ} < |\phi| < 180^{\circ}$, 500 m $< R <$ 4500 m. 
The red, green, blue, yellow, magenta and black represent gamma, electron, muon, other shower components, atmospheric muon background and the total of them, respectively.
The vertical error bar shows the standard deviation. 
(bottom) Lateral distributions of the muon purity. 
The violet and orange show calculations with and without the atmospheric muon background, respectively.
\label{fig:cLDF_2}}
\end{figure}

The statistical error of the average signal cannot be simply calculated for $R \gtrsim 1500$~m.
It is because the average number of air shower particles is less than unity in the region so the fraction of SDs with no hit signals is too large to determine lower and upper errors from the shape of the signal size distribution.
Hence we assume the Poisson distribution {$f(x) = N^x e^{-N} / x!$; $N$ is the average value of the distribution and $x$ is the variable} for the signal size distribution.
We calculated the average signal by the following equation:
%\begin{equation}
$n_0 / n_{\rm{all}} = f(0) = e^{-N}$.
%\end{equation}
Here $n_0$ and $n_{\rm{all}}$ are the entries of 0 VEM bin and the whole distribution, respectively.
The probability that zero values appear $n_0$ times in $n_{\rm{all}}$ samples follows the binomial distribution, hence the standard deviation of $n_0$ is calculated as $\sqrt{n_{\rm{all}}\, p(1-p)}$, where $p$ is $n_0 / n_{\rm{all}}$.
Using these considerations, we calculate the average signal $N$ and the statistical error.

\section{Results} \label{sec:5}

\subsection{Comparison of data with MC}
Figure \ref{fig:pLDF_1} shows the lateral distributions of the signal and the ratio of the data to proton MC using the QGSJET I$\hspace{-.1em}$I-03 hadronic interaction model.
The average ratios of the data to the MC are calculated to be $1.72\pm0.10{\rm (stat.)}\pm0.37{\rm (syst.)}$ at 1910~m~$<~R~<$~2160~m and $3.14 \pm 0.36{\rm (stat.)} \pm 0.69 {\rm (syst.)}$ at 2760~m~$< R <$~3120~m.
The systematic uncertainty is explained in Section \ref{sec:syst}.
The observed lateral distribution falls down slower than the MC.
The data becomes closer to the MC at $R~\gtrsim~4000$~m,
since the atmospheric muon background dominates the SD signals at the distance.

Figure \ref{fig:pLDF_2} shows the lateral distributions of the signal and the ratio of the data to the MC with other hadronic interaction models; QGSJET I$\hspace{-.1em}$I-03, QGSJET I$\hspace{-.1em}$I-04, Epos 1.99 and Sibyll 2.1.
The ratios of the data to the MC with QGSJET I$\hspace{-.1em}$I-04 for proton are $1.67 \pm 0.10{\rm (stat.)} \pm 0.36 {\rm (syst.)}$ at 1910~m~$< R <$~2160~m and $2.75 \pm 0.32{\rm (stat.)} \pm 0.60 {\rm (syst.)}$ at 2760~m~$< R <$~3120~m.
The observed lateral distribution (circles) decreases less with radial distance than that of all hadronic interaction models (other points).

\begin{figure}
   \includegraphics[width=0.45\textwidth] {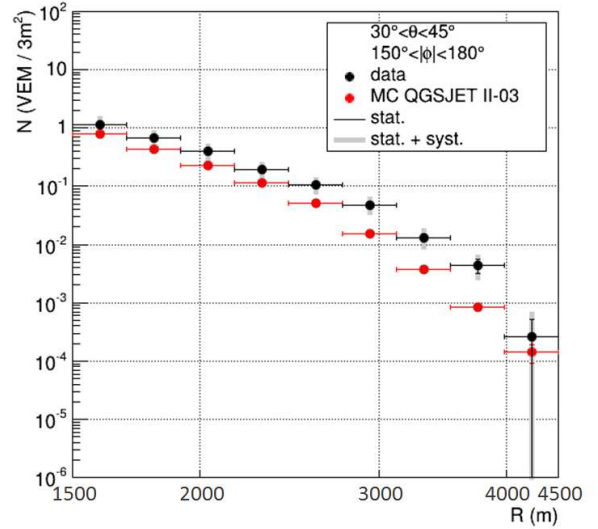}
   \includegraphics[width=0.45\textwidth] {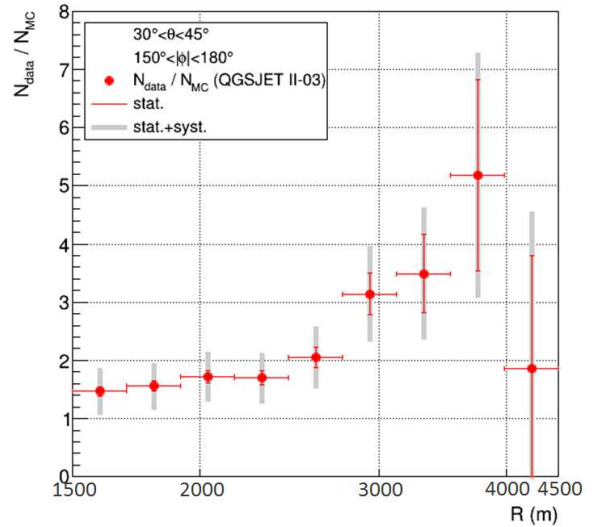}
\caption{(color online).
Lateral distributions of air showers of the data and the MC for $30^{\circ} < \theta < 45^{\circ}, 150^{\circ} < |\phi| < 180^{\circ}$, 1500 m $< R <$ 4500 m. 
The vertical thin error bars and shaded grey thick error bars represent statistical errors and quadratic sums of statistical and systematic errors, respectively.
(top) Lateral distributions of the average signal assuming the histograms follow the Poisson distribution, denoted as $N$ in the figure. 
The black and red points represent the data and the MC, respectively. 
(bottom) The average ratio of the data to the MC. 
\label{fig:pLDF_1}}
\end{figure}

\begin{figure}
   \includegraphics[width=0.45\textwidth] {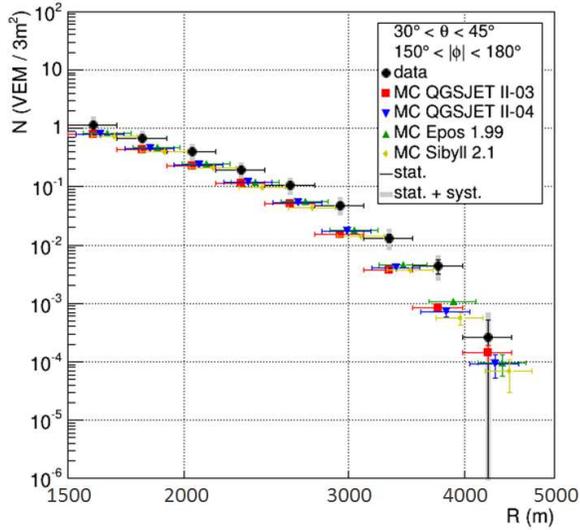}
   \includegraphics[width=0.45\textwidth] {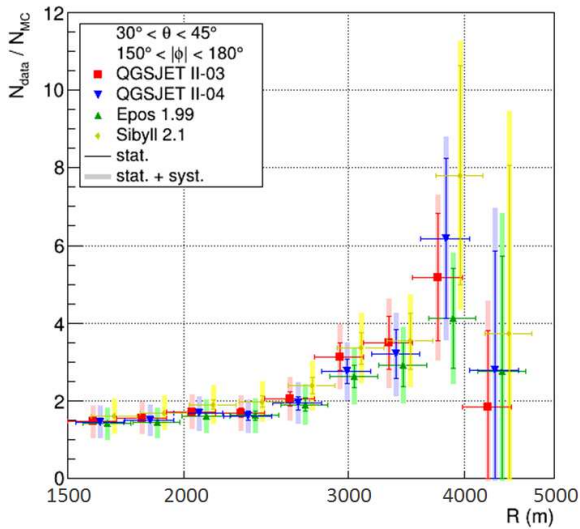}
\caption{(color online).
Same as Figure \ref{fig:pLDF_1}, but with the MCs using other hadronic models. 
(top) Lateral distributions of the average signal size assuming the histograms follow the Poisson distribution. 
The black, red, blue, green and yellow represent data, QGSJET I$\hspace{-.1em}$I-03, QGSJET I$\hspace{-.1em}$I-04, Epos 1.99 and Sibyll 2.1, respectively.
To make error bars easy to see, the plots for the latter three models are shifted to the right. 
(bottom) The average ratio of the data to the MC. 
The color corresponds to MC hadronic models described in the top figure.
\label{fig:pLDF_2}}
\end{figure}

We calculated lateral distributions for iron showers using the MC with QGSJET I$\hspace{-.1em}$I-03.
Figure \ref{fig:iron} shows lateral distributions of the ratio of the data to the MCs for proton and iron.
The average signal of the data is larger than the MC for iron for $R \gtrsim 2500$~m.
For the smaller distances, the difference between the data and the MC for iron is smaller than systematic errors.
Table \ref{tab:res} summarizes the results in each $R$.

\begin{table}
  \begin{ruledtabular}
    \caption{Ratio of observed SD signal sizes to MC predictions using QGSJET I$\hspace{-.1em}$I-03 as a function of $R$. 
Errors due to statistical error ($stat.$) and systematic error ($syst.$) are described. \label{tab:res}}
    \begin{tabular}{l c c} 
\,      $R$ [m] & Ratio $\pm \sigma_{\rm stat.} \pm \sigma_{\rm syst.}$ &  \\ 
       & Proton & Iron \\ \hline 
\,      [1500, 1695] & $1.47 ^{+0.09}_{-0.08} \pm 0.35$ &  $ 1.16 ^{+0.07}_{-0.06} \pm 0.28 $ \\ 
\,      [1695, 1915] & $ 1.56 ^{+0.09}_{-0.08} \pm 0.35 $ &  $ 1.16 \pm 0.06 \pm 0.26 $ \\ 
\,      [1915, 2160] & $1.72 \pm 0.10 \pm 0.37$ &  $ 1.26 \pm 0.07 \pm 0.27 $ \\ 
\,      [2160, 2445] & $ 1.69 \pm 0.12 \pm 0.37  $ & $ 1.22 \pm 0.08 \pm 0.27$ \\ 
\,      [2445, 2760] &  $ 2.05 \pm 0.18 \pm 0.46 $ & $ 1.38 \pm 0.11 \pm 0.31 $ \\ 
\,      [2760, 3120] &  $ 3.14 \pm 0.36 \pm 0.69$ & $ 1.74 \pm 0.19 \pm 0.38 $ \\ 
\,      [3120, 3525] & $ 3.49 \pm 0.68 \pm 0.86 $ & $ 1.71 \pm 0.30 \pm 0.42 $ \\ 
\,      [3525, 4180] & $ 5.18 \pm 1.64 \pm 1.27 $ & $ 2.96 \pm 0.83 \pm 0.72 $ \\ 
\,      [4180, 4500] & $ 1.85 \pm 1.95 \pm 1.81 $ & $ 0.99 \pm 0.99 \pm 0.96 $ \\
    \end{tabular}
  \end{ruledtabular}
\end{table}

Figure \ref{fig:mu_dm2} shows the correlation between the muon purity expected from the MC and the ratio of the signal size of the data to that of the MC. 
We loosened the cut condition of the zenith angle of air showers from $45^{\circ}$ to $55^{\circ}$ to see the correlation precisely.
On the high muon purity condition (30$^{\circ}~<~\theta~<~45^{\circ}, 150^{\circ}~<~|\phi|~<~180^{\circ}$, 2000~m $< R <$ 4000~m, magenta filled circle in Figure \ref{fig:mu_dm2}), the muon purity and the ratio of the data to the MC are 65\% and $1.88 \pm 0.08{\rm (stat.)} \pm 0.42 {\rm (syst.)}$, respectively.
In the case of the low muon purity condition ($\theta < 30^{\circ}, |\phi| < 30^{\circ}, 2000$~m $< R <$ 4000~m, black open circle in Figure \ref{fig:mu_dm2}), 
they are calculated to be 28\% and $1.30 \pm 0.06{\rm (stat.)} \pm 0.29 {\rm (syst.)}$, respectively.
This figure shows larger differences in signal sizes between the data and the MC for conditions of higher muon purity.

\begin{figure}[tb]
   \includegraphics[width=0.45\textwidth] {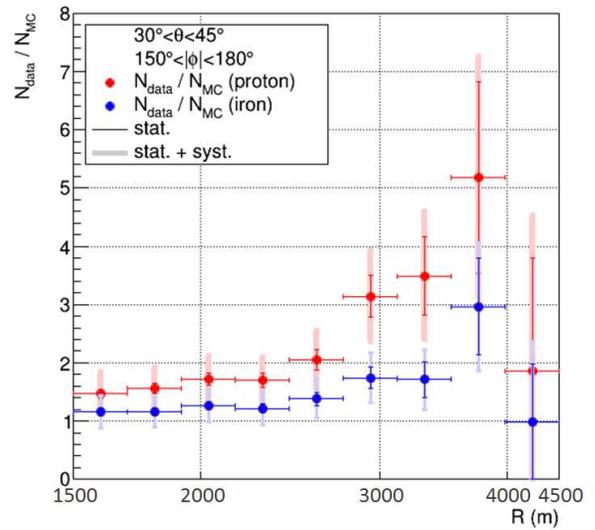}
\caption{(color online).
Ratios of the signal size of the data to the MCs for proton and iron. 
The vertical thin error bars and shaded thick error bars represent statistical errors and quadratic sums of statistical and systematic errors, respectively.
The red and blue points represent the ratios of the data to the MC for proton and that for iron, respectively.
\label{fig:iron}}
\end{figure}

\begin{figure}[tb]
   \includegraphics[width=0.45\textwidth] {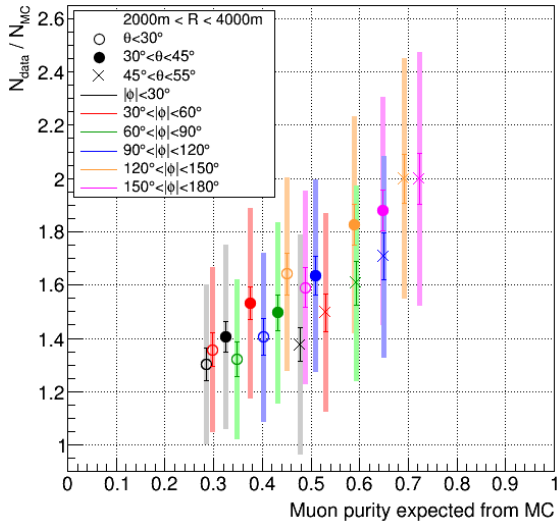}
\caption{(color online).
The correlation between the muon purity and the ratio of the signal size of the data to the MC with QGSJET I$\hspace{-.1em}$I-03 for 2000 m $< R <$ 4000 m. 
The black, red, green, blue, orange and magenta represent $|\phi| < 30^{\circ}$, $30^{\circ} < |\phi| < 60^{\circ}$, $60^{\circ} < |\phi| < 90^{\circ}$, $90^{\circ} < |\phi| < 120^{\circ}$, $120^{\circ} < |\phi| < 150^{\circ}$ and $150^{\circ} < |\phi| < 180^{\circ}$, respectively. 
The open circle, filled circle and cross represent $\theta < 30^{\circ}$, $30^{\circ} < \theta < 45^{\circ}$ and $45^{\circ} < \theta < 55^{\circ}$, respectively.
The vertical thin error bars and shaded thick error bars represent statistical errors and quadratic sums of statistical and systematic errors, respectively.
\label{fig:mu_dm2}}
\end{figure}

\subsection{Systematic uncertainty} \label{sec:syst}

$FD \,\, energy \,\, determination$: One of the systematic uncertainties of this work is caused by the uncertainty of the TA FD energy measurement, which is 21\% \cite{BRLRhybrid3}.
According to the generalized Heitler model of hadronic air showers, the number of particles from the EM and muonic components of the showers are proportional to $E^{1.03}$ and $E^{0.85}$ respectively, where $E$ is the primary cosmic ray energy \cite {heit}.
In this analysis, SD signals include both EM and muon components, with a muon fraction of 60-70\%.
We conservatively assume the signal size is proportional to E and apply the systematic uncertainty of $\pm 21$\% to the measurement.

$1\,\,MIP\,\,calibration$: 
1 MIP signal size is determined by fitting a histogram of single atmospheric muons and searching for the peak position of the histogram.
The accuracy of this calibration method is calculated as $\epsilon_{\rm 1mip} / S_{\rm 1mip}$, where $S_{\rm 1mip}$ is the peak value of the histogram determined by fitting and $\epsilon_{\rm 1mip}$ is the fitting error of $S_{\rm 1mip}$.
We calculate the average value of $\epsilon_{\rm 1mip} / S_{\rm 1mip}$ using all the SD signals in the dataset to estimate the systematic error.
The calculated error value is $\pm 1.2\%$.

$Atmospheric\,\,muon\,\,cut$:
In the TA SD event reconstruction, 
we perform a cut on the SD signals not included in space-time clusters.
This procedure reduces random atmospheric muon background in the dataset.
We calculated the systematic error of this procedure as the difference in the cut signal ratio between the data and the MC, that is $\pm |(S_{\rm cut} / S_{\rm no \,cut})_{\rm data} - (S_{\rm cut} / S_{\rm no \,cut})_{\rm MC}| / 2$.
Here $S_{\rm cut}$ and $S_{\rm no \,cut}$ are the signal before and after the cut, respectively.
To avoid air shower signals affecting the calculation, we used
the bin at 4000~m $\lesssim R <$ 4500~m.
The background in that bin is expected to be larger than each air shower component from the MC.
The calculated error is $\pm 1\%$.

$Poisson\,\,distribution\,\,assumption$:
In this analysis, we calculated the average signal from air showers with an assumption of the signal size distribution following the Poisson distribution.
It is possible that these distributions do not match due to a smearing effect of signals in the SD.
We calculate the systematic error from this assumption by comparing the average signal using the Poisson distribution, $N$, with the simple averaged value, $S$;
$\pm |(S_{\rm data} / S_{\rm MC} - N_{\rm data} / N_{\rm MC}) / (N_{\rm data} / N_{\rm MC})| / 2$.
The calculated values have $R$ dependence and are within $\pm 4 \%$.

$Event\,\,reconstruction$:
The TA SD reconstruction procedure uses the SDs in the lateral distribution fitting without separating them by the azimuth angle.
The signal size of air showers in the shower arrival direction is larger than that in the shower forwarding direction compared in same $R$, thus the reconstructed core position has a systematic shift on the side of the air shower arrival direction.
If the shift of the data is different from that of the MC, it will result in a systematic error. 
However, we cannot compare the shift between the data and the MC since the ``true'' air shower geometry of the data is not given.
Instead we calculated the bias of the signal size from the shift, that is $\pm (N_{\rm in} - N_{\rm rec}) / N_{\rm rec} / 2$, using the MC.
Here $N_{\rm{rec}}$ and $N_{\rm{in}}$ are the signal size with reconstructed event parameters ($E, \theta, \phi$ and core position) and same with input ones, respectively.
The calculated values have $R$ dependence and are in the range of 4 - 13\%.
We used them as the systematic error.

$SDs\,\,not\,working\,properly$:
The average duty cycle of the SD array is approximately 95\%, hence 5\% of SDs in the event dataset may be assumed to be not working properly. 
We calculated average signal sizes with this effect removed and compared it with the measured value.
The systematic uncertainty is calculated by subtracting this bias between the data and the MC.
The calculated values have $R$ dependence and are within $\pm 1 \%$.

Table \ref{tab:tot_syst} summarizes all systematic errors in this work. 
As a result of above considerations, the energy determination uncertainty dominates the total systematic errors.

\begin{table}[]
  \begin{ruledtabular}
    \caption{Summary of systematic errors in the TA SD signal on the condition $30^{\circ} < \theta < 45^{\circ}$, $150^{\circ} < |\phi| < 180^{\circ}$ and 2000~m $< R <$ 4000~m.\label{tab:tot_syst}}
    \begin{tabular}{l c} 
      Source & Systematic error \\ \hline 
      FD energy determination & $\pm$21\% \\ 
      1 MIP calibration & $\pm$1.2\% \\ 
      Atmospheric muon cut & $\pm$1\% \\ 
      Poisson distribution assumption & $\pm(<4$\%) \\ 
      Event reconstruction & $\pm$(4-13\%) \\ 
      SDs not working properly & $\pm(<1$\%)  \\ \hline
      Total & $\pm$(22-24\%) \\ 
    \end{tabular}
  \end{ruledtabular}
\end{table}

\section{Discussion} \label{sec:6}
In this paper, we have established a method to study muons from air showers with the TA scintillator SD.
The results imply that part of the discrepancy in signal sizes between the data and the MC is due to a muon excess in the data.
The measurement presented here is qualitatively consistent with the result of the Auger experiment \cite{Au_Nmu1, Au_Nmu2},
in which the threshold energy of muon detection ($\sim$10~MeV for TA and 300~MeV for Auger) and $R$ conditions are different.
We also found that a lateral distribution in the data falls slower than that of the MC on a high muon purity condition.
It suggests the lateral distribution of muons differs between data and MCs.

Another lesson from this analysis is that the muon excess is part of the cause of an energy scale discrepancy between the TA SD and FD.
Comparing the energy measured by the SD and that by the FD in the same events, we see a difference of 27\%  \cite{TAspe}.
The TA FD measures an air shower energy from calorimetric fluorescence light signals, which are mostly from the EM component and less affected by the excess of muons.
We use $S800$ to determine the energy with the TA SD, where $S800$ is the charge density measured at 800 m from the shower axis. 
Though the muon purity in SD signals is about 20\% at that distance, the muon excess may partly cause misunderstanding of the conversion relation between $S800$ and the primary energy in the MC, resulting in the systematic shift of the measured energy scale.

\section{Conclusion} \label{sec:7}

We have studied muons in UHECR-induced air showers detected by the TA SD.
The air shower events and the locations of SDs were binned in $\theta, \phi$ and $R$ in order to determine events with a high muon purity.
We compared air shower signals of the data with that of the MC on the high muon purity condition ($30^{\circ} < \theta < 45^{\circ}, 150^{\circ} < |\phi| < 180^{\circ}$, 2000~m $< R <$ 4000~m) for $10^{18.8}$~eV $< E <$ $10^{19.2}$~eV.
On that condition, the muon purity expected from the MC is $\sim$65\%, and the ratios of the signal size of the data to that of the MC with QGSJET I$\hspace{-.1em}$I-03 are $1.72 \pm 0.10{\rm (stat.)} \pm 0.37 {\rm (syst.)}$ at 1910~m $< R <$ 2160~m and $3.14 \pm 0.36{\rm (stat.)} \pm 0.69 {\rm (syst.)}$ at 2760~m $< R <$ 3120~m.
The ratios for the MC with QGSJET I$\hspace{-.1em}$I-04 are $1.67 \pm 0.10{\rm (stat.)} \pm 0.36 {\rm (syst.)}$ and $2.75 \pm 0.32{\rm (stat.)} \pm 0.60 {\rm (syst.)}$ at the same $R$ bins.
The lateral distribution of the data falls down slower than the MC on the high muon purity condition, resulting in a larger ratio of the data to MC at larger $R$ values.
Also, the difference in the signal size between the data and the MC is larger with higher muon purity, which implies that part of the discrepancy between the data and the MC is due to an excess of muons in the data. 

The primary effect found in this work, that the muon signal is larger in the data than the MC, is qualitatively consistent with the excesses of muons observed in the Pierre Auger Observatory. 
In addition, we found that the shape of the lateral distribution of the MC did not reproduce that found in the data. 
This result provides information critical to the understanding of hadronic interactions at ultra-high energies, and to improving the reliability of hadronic interaction models used in the air shower MC.

\begin{acknowledgments}

The Telescope Array experiment is supported by the Japan Society for the Promotion of Science (JSPS) through Grants-in-Aid for Priority Area 431, for Specially Promoted Research JP21000002, for Scientific Research (S) JP19104006, for Specially Promote Research JP15H05693, for Scientific Research (S) JP15H05741 and for Young Scientists (A) JPH26707011; by the joint research program of the Institute for Cosmic Ray Research (ICRR), The University of Tokyo; by the U.S. National Science Foundation awards PHY-0601915, PHY-1404495, PHY-1404502, and PHY-1607727; by the National Research Foundation of Korea (2015R1A2A1A01006870, 2015R1A2A1A15055344, 2016R1A5A1013277, 2016R1A2B4014967, 2017R1A2A1A05071429); by the Russian Academy of Sciences, RFBR grant 16-02-00962a (INR), IISN project No. 4.4502.13, and Belgian Science Policy under IUAP VII/37 (ULB). The foundations of Dr. Ezekiel R. and Edna Wattis Dumke, Willard L. Eccles, and George S. and Dolores Dore Eccles all helped with generous donations. The State of Utah supported the project through its Economic Development Board, and the University of Utah through the Office of the Vice President for Research. The experimental site became available through the cooperation of the Utah School and Institutional Trust Lands Administration (SITLA), U.S. Bureau of Land Management (BLM), and the U.S. Air Force.  We appreciate the assistance of the State of Utah and Fillmore offices of the BLM in crafting the Plan of Development for the site. Patrick Shea assisted the collaboration with valuable advice on a variety of topics. The people and the officials of Millard County, Utah have been a source of steadfast and warm support for our work which we greatly appreciate. We are indebted to the Millard County Road Department for their efforts to maintain and clear the roads which get us to our sites. We gratefully acknowledge the contribution from the technical staffs of our home institutions. An allocation of computer time from the Center for High Performance Computing at the University of Utah is gratefully acknowledged.

\end{acknowledgments}


\begin{thebibliography}{99}
\bibitem{TA} M. Fukushima \emph{et al.}, 
Prog. Theor. Phys. Suppl. {\bf 151}, 206 (2003).

\bibitem{Xmax_stereo} T. Stroman \emph{et al.}, 
Proc. 34th ICRC, 361 (2015).

\bibitem{Xmax_mono} T. Fujii \emph{et al.}, 
Proc. 34th ICRC, 320 (2015).

\bibitem{hir} T. Abu-Zayyad \emph{et al.}, 
Phys. Rev. Lett. {\bf 84}, 4276 (2000).

\bibitem{yak} A. V. Glushkov \emph{et al.}, 
JETP Lett. {\bf 87}, 190 (2008).

\bibitem{Au_Nmu1} A. Aab \emph{et al.}, 
Phys. Rev. D {\bf 91}, 032003 (2015); Erratum, Phys. Rev. D {\bf 91}, 059901 (2015).

\bibitem{QII-03} S. Ostapchenko, 
Nucl. Phys. B, Proc. Suppl. {\bf 151}, 147 (2006).

\bibitem{QII-04} S. Ostapchenko, 
Phys. Rev. D {\bf 83}, 014018 (2011).

\bibitem{EP-LHC} K. Werner, F. M. Liu, and T. Pierog, 
Phys. Rev. C {\bf 74}, 044902 (2006).

\bibitem{Au_Nmu2} A. Aab \emph{et al.}, 
Phys. Rev. Lett. {\bf 117}, 192001 (2016).

\bibitem{kas} W.D. Apel \emph{et al.}, 
Astropart. Phys. {\bf 95}, 25 (2017).

\bibitem{atm} L.G. Dedenko \emph{et al.}, 
EPJ Web Conf. {\bf 99}, 10003 (2015).

\bibitem{ice} JG Gonzalez \emph{et al.}, 
J. Phys. Conf. Ser. {\bf 718}, 052017 (2016).

\bibitem{eas} Yu.A. Fomin \emph{et al.}, 
Astropart. Phys. {\bf 92}, 1 (2017).

\bibitem{SD2012} T. Abu-zayyad \emph {et al.}, 
Nucl. Instrum. Methods. A {\bf 689}, 87 (2012).

\bibitem{FD_BRLR} R.U. Abbasi \emph{et al.}, 
Astropart. Phys. {\bf 80}, 131 (2016).

\bibitem{DI} D. Ivanov, 
Ph.D. thesis, Rutgers-The State University of New Jersey, Department of Physics and Astronomy, Piscataway, New Jersey, USA (2012).

\bibitem{lat1} M. Takeda \emph{et al.}, 
Phys. Rev. Lett. {\bf 81}, 1163 (1998).

\bibitem{lat2} M. Takeda \emph{et al.}, 
Astropart. Phys. {\bf 19}, 447 (2003).

\bibitem{CORSIKA} D. Heck \emph{et al.}, 
Forschungszentrum Karlsruhe Report FZKA, 6019 (1998).

\bibitem{Xmax_BRLRhybrid2} D. Ikeda \emph{et al.}, 
Proc. 34th ICRC 362 (2015).

\bibitem{BRLRhybrid3} T. Abu-Zayyad \emph{et al.}, 
Astropart. Phys. {\bf 61}, 93 (2015).

\bibitem{DI_2017} R. U. Abbasi \emph{et al.}, 
Astropart. Phys. {\bf 86}, 21 (2017).

\bibitem{flu1} A. Ferrari \emph{et al.}, 
Tech. Rep. 2005-010, CERN (2005).

\bibitem{flu2} G. Battistoni \emph{et al.}, 
AIP Conf. Proc. {\bf 896}, 31 (2007).

\bibitem{egs4} W. R. Nerson \emph{et al.}, 
Tech. Rep. 0265 SLAC (1985).

\bibitem{thin} M. Kobal \emph{et al.}, 
Astropart.Phys. {\bf 15}, 259 (2001).

\bibitem{dethin} B. T. Stokes \emph{et al.}, 
Astropart.Phys. {\bf 35}, 759 (2012).

\bibitem{Geant4} J. Allison \emph{et al.}, 
IEEE Trans. Nucl. Sci. {\bf 53}, 270 (2006).

\bibitem{hires} R. U. Abbasi \emph{et al.}, 
Phys. Rev. Lett. {\bf 100}, 101101 (2008).

\bibitem{TAspe} T. Abu-Zayyad \emph{et al.}, 
ApJ {\bf 768}, L1 (2013).

\bibitem{Xmax_MDhybrid2} R.U. Abbasi \emph{et al.}, 
Astropart. Phys. {\bf 64}, 49 (2015).

\bibitem{EP199} T. Pierog and K. Werner, 
Nucl. Phys. Proc. Suppl. {\bf 196}, 102 (2009).

\bibitem{Sib2.1} E. J. Ahn \emph{et al.}, 
Phys. Rev. D {\bf 80}, 094003 (2009).

\bibitem{heit} J. Matthews, 
Astropart. Phys. {\bf 22}, 387 (2005).

\end{thebibliography}
\end{document}